\documentclass[preprint,12pt]{elsarticle}

\usepackage[a4paper,bindingoffset=0.2in,%
            left=1in,right=1in,top=1.5in,bottom=1.5in,%
            footskip=.25in]{geometry}
            
\usepackage{amssymb}

\usepackage{hyperref}
\usepackage{amsmath}
\usepackage{float}

\begin{document}

\begin{frontmatter}



\title{Compelling new electrocardiographic markers for automatic diagnosis}


\author[a]{Cristina Rueda\corref{cor1}}
\ead{cristina.rueda@uva.es}
\author[a]{Itziar Fern\'andez} 
\author[a]{Yolanda Larriba}
\author[a]{Alejandro Rodr\'iguez-Collado}
\author[a]{Christian Canedo}

\address[a]{Department of Statistics and Operations Research, Universidad de Valladolid, Paseo de Bel\'en 7, 47011, Valladolid, Spain}
\cortext[cor1]{Corresponding author}

\begin{abstract}
\textit{Background and Objective:} The automatic diagnosis of  heart diseases from the electrocardiogram (ECG) signal is crucial in clinical decision-making. However, the use of computer-based decision rules in clinical practice is still deficient, mainly due to their complexity and a lack of medical interpretation. The objetive of this research is to address these issues by providing valuable diagnostic rules that can be easily implemented in clinical practice. In this research, efficient diagnostic rules friendly in clinical practice are provided. \textit{Methods:} In this paper, interesting parameters obtained from the ECG signals analysis are presented and two simple rules for automatic diagnosis of Bundle Branch Blocks are defined using new markers derived from the so-called FMM$_{ecg}$ delineator. The main advantages of these markers are the good statistical properties and their clear interpretation in clinically meaningful terms. \textit{Results:} High sensitivity and specificity values have been obtained using the proposed rules with data from more than 35000 patients from well known benchmarking databases. In particular, to identify Complete Left Bundle Branch Blocks and differentiate this condition from subjects without heart diseases, sensitivity and specificity values ranging from 93\% to 99\% and from 96\% to 99\%, respectively. The new markers and the automatic diagnosis are easily available at \url{https://fmmmodel.shinyapps.io/fmmEcg/}, an \textit{app} specifically developed for any given ECG signal. \textit{Conclusions:} The proposal is different from others in the literature and it is compelling for three main reasons. On the one hand, the markers have a concise electrophysiological interpretation. On the other hand, the diagnosis rules have a very high accuracy. Finally, the markers can be provided by any device that registers the ECG signal and the automatic diagnosis is made straightforwardly, in contrast to the black-box and deep learning algorithms.
\end{abstract}


\end{frontmatter}


\section{Introduction}
\label{S:1}
The relevance of the automatic diagnosis of diseases is beyond the debate. In the case of cardiovascular diseases, it can be a matter of life and death. Many researchers have developed automatic rules for diagnoses, but their use in clinical practice is still deficient. One reason is that many cases are based on black-box algorithms that physicians do not trust because they lack a medical interpretation. Furthermore, the validation process is often not clean, as it depends heavily on the selection of signals and patients (training and test) and on the preprocessing stage. In addition to these limitations of most studies, there are those derived for the lack of consensus on the definition of electrocardiographic features and criteria to diagnose certain diseases using such features.

\begin{figure}[h!]
   \centering
  \includegraphics[width=1\textwidth]{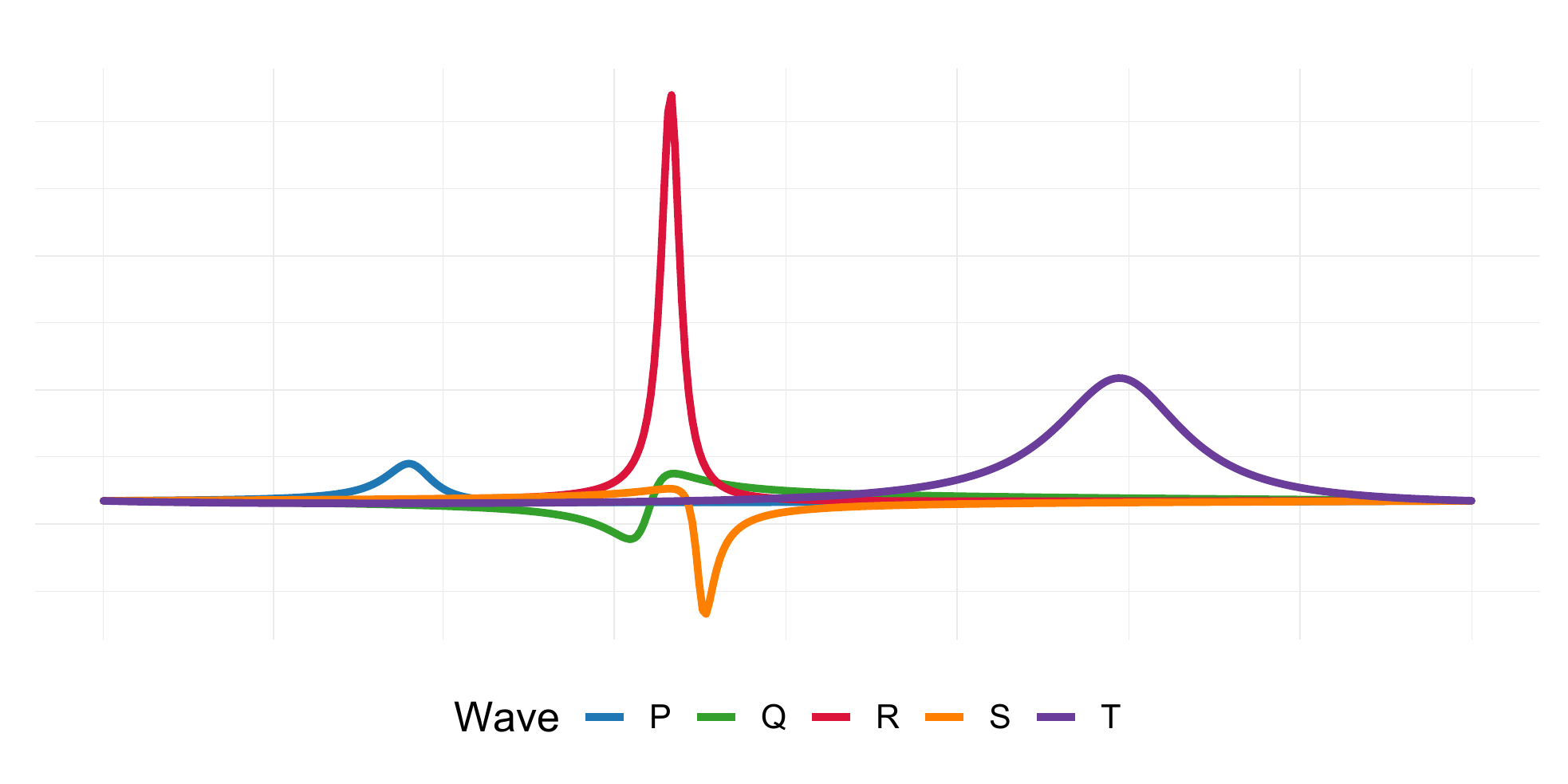}
  \caption{The five waves: P, Q, R, S, T identify by the FMM$_{ecg}$ approach for a typical non-pathological pattern. Data from patient $37$ from the Georgia database.}
\label{f:typ}
\end{figure}

The proposal in this paper is to prove that we can get around all these problems by the definition of new markers using the FMM$_{ecg}$ delineator \cite{Rueda+Larriba+Lamela:2021}. 
The FMM$_{ecg}$ delineator describes the fragment of an ECG signal corresponding to a heartbeat as the combination of five waves corresponding to the fundamental waves in a heartbeat: P, Q, R, S, T. The details of the model specifications, such as  the restrictions and the estimation algorithm are given in \cite{Rueda+Larriba+Lamela:2021}. In addition, that paper reveals the methodology's potential to describe a variety of non-pathological, noisy, and pathological ECG patterns. The FMM$_{ecg}$ wave decomposition for a typical non-pathological pattern, from a left lead signal, is shown in Figure \ref{f:typ}. 

In particular, using the FMM$_{ecg}$ model parameters, two  powerful markers, \textit{omeR} and \textit{omeS}, are presented here and are used as rules for the diagnosis of the Bundle Branch Block (BBB), which is a defect in the electrical conduction system of the heart. In such cases, ventricular enlargement or hypertrophy occurs, and the QRS complex will widen, deform and prolong \cite{McAnulty+Rahimtoola:1984}. 

On the one hand, a left BBB (LBBB) happens when the activation of the left ventricle of the heart is delayed, which causes the left ventricle to contract later than the right ventricle. In particular, LBBB is associated to a higher risk of suffering different cardiovascular diseases, including a high mortality ascribed to acute myocardial infarction. Some authors also noted the crucial need for a prompt identification of LBBB in determining the best health strategy and future for patients. Specifically relevant is the detection of patients with complete LBBB (CLBBB) who may respond positively to cardiac resynchronization therapy. \cite{Tan+etal:2020} is a good review of the state of the art in the LBBB disease. 

On the other hand, the right BBB (RBBB) is a blockage of electrical impulses to the heart’s right ventricle and it is one of the most common electrocardiographic abnormalities that is often detected in asymptomatic patients \cite{DeBacquer+DeBacquer+Kornitzer:2000}. Although the RBBB has been associated with fewer complications for cardiovascular disease in comparison to LBBB \cite{Breithardt+Breithardt:2012}, the complete RBBB (CRBBB) has an impact on patients with other diseases, including acute myocardial infarction, and the appearance of CRBBB in patients with other cardiovascular diagnoses worsen their prognosis \cite{Alventosa+etal:2019}, so it is advisable to develop tools for early detection.

\begin{figure*}[!t]
	\centering
	\includegraphics[width=1\textwidth]{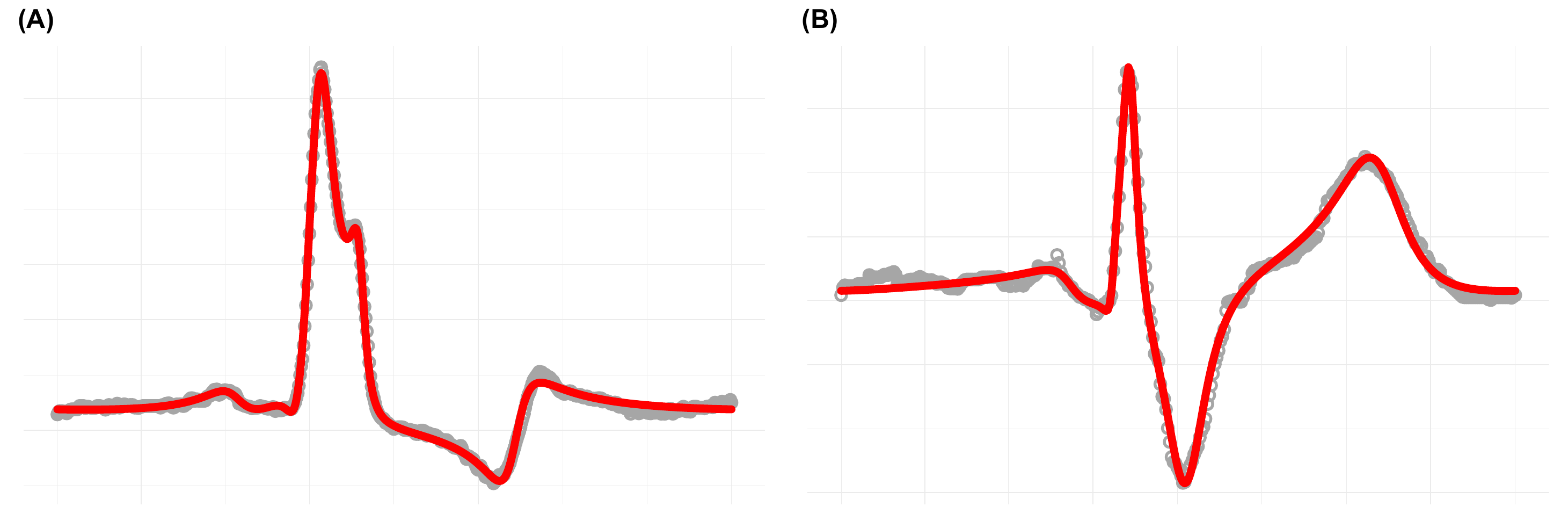}
	\caption{Typical ECG morphology in complete BBB. ECG signals are represented as grey points and FMM$_{ecg}$ predictions as a red line. Left panel (A): data from patient 2087 (CPCS database) with CLBBB diagnosis.  Right panel (B): data from patient 10131 (PTB-XL database) with CRBBB diagnosis.}
	\label{f:BBBPatterns}
\end{figure*}

The electrographic diagnosis criteria for CLBBB and CRBBB depends on the location of the lead analyzed. In this paper, we initially consider only left-sided leads, specifically: I, II and V5. 
The typical ECG signals, in the presence of complete BBB (CBBB), are shown in Figure~\ref{f:BBBPatterns} for lead I. In agreement with the figure, the  widely mentioned CLBBB diagnostic criteria for such leads can be summarized in four points as follows \cite{Willems+etal:2000,Surawicz+etal:2009,Strauss+Selvester+Wagner:2011, Tan+etal:2020}:

\begin{enumerate}
\item Prolonged QRS complex duration. Even longer for male than female.
\item Notched and slurred R-waves. 
\item Prolonged R wave peak time.
\item T wave inversions and ST-segment depression.
\end{enumerate}

It worth mentioning the Strauss criteria \cite{Strauss+Selvester+Wagner:2011} as it is widely used and closely related to the criteria recommended by different institutions as the American Heart Association Electrocardiography and Arrhythmias Committee, the Council on Clinical Cardiology/American College of Cardiology Foundation/Heart Rhythm Society~\cite{Surawicz+etal:2009}. The Strauss criteria is based on the first two points in the enumeration above.

Concerning CRBBB, although there is no consensus in its diagnosis criteria \cite{Alventosa+etal:2019}, most researchers assume the following points as typical CRBBB signs for left-sided leads:
\begin{enumerate}
\item Large QRS complex duration.
\item Wide, negative, and slurred S wave.
\item Large S wave duration.
\end{enumerate}

The ECG findings described above for CLBBB or CRBBB diagnoses are apparently simple but nothing is further from the truth. First, there is no consensus on the definition of the QRS complex duration or the R wave peak time. The T wave inversion or the ST-segment depression are also vague descriptions \cite{Kligfield+etal:2018}. There is no standard definition of the QRS notch and slur patterns \cite{Xia+etal:2016}, and no simple criteria exist to determine their number and location. Moreover, many of the well-known ECG features, such as the QRS complex duration, strongly depend on the heart rate \cite{Mason+etal:2016}. In particular, this association confounds the relationship between QRS durations and sex, which affects the definition of the criteria for diagnosis. In addition, a considerable QRS duration is a typical characteristic in both CLBBB and CRBBB which makes the difference a problematic task.

Finally, significant systematic differences among engineering solutions of manufacturers of automated ECG have been detected due to preprocessing and denoising stages and the algorithms used. Specifically, such points as the end of the R wave and the end of the QRS complex have no precise definition. Consequently, different results might be expected of the QRS duration and other features of the same underlying ECG waveform by different algorithms \cite{Kligfield+etal:2018}.
 
In the last few years, together with a considerable increase in electronic health databases, such as  Physionet's Physiobank \cite{Goldberger+etal:2000}, machine learning algorithms, including deep learning, have become popular in the automatic diagnosis of cardiac abnormalities, see \cite{Lyon+etal:2018,Hannun+etal:2019,Ribeiro+etal:2020} among others. In particular, few studies have focused exclusively on the automated identification of BBB, we have found  \cite{Hao+etal:2018} and \cite{Allami+etal:2016}. In those cases, the classifiers achieved sensitivity and specificity values above 95\% for RBBB and 98\% for LBBB. However, these algorithms have been developed and tested using ECG recordings  from only $47$ subjects from the MIT-BIH Arrhythmia database \cite{Moody+Mark:2001}.  Furthermore, there has been recently  a proliferation of methods for the automatic diagnosis of CLBBB patients known to have a good clinical response to cardiac resynchronization therapy. For example, \cite{MartinYebra+Martinez:2019} proposed a method based on wavelet analysis, obtaining high and moderate values of sensitivity and specificity (92.9\% and 65.1\%, respectively). A more accurate diagnosis was achieved in \cite{Yang+etal:2020}, which combines a random forest classifier and a neural network with sensitivity and specificity values of 91.7\% and 88.7\%, respectively. Both algorithms were trained and validated using data on 600 patients from the Multicenter Automatic Defibrillator Implantation Trial—Cardiac Resynchronization Therapy (MADIT-CRT) database \cite{Moss+etal:2009}. It is important to note that these classifiers use the same database for training, test and validation and that the number of patients is limited, which increases the risk of overfitting and failure when used in any other database.  Another well-known drawback of machine learning is the lack of any physiological interpretation of the rules. These limitations have prevented its use in clinical practice. 

The new markers, $omeR$ and $omeS$, presented here have a consistent mathematical and morphological meaning, with a definition independently of the electrocardiograph or expert, and this consequently makes them universal and quite useful in diagnosis. In particular, \textit{omeR}  captures the width of the R wave, is sex-independent, is closely related to the duration of QRS complex and the R peak time, and emerges as a measure of the degree of LBBB severity. Moreover, \textit{omeS} captures the relevance of negative S waves. A complementary rule combining \textit{omeR} and \textit{omeS} is useful to diagnose CBBB. 

The potential of new markers and the corresponding diagnostic rules are validated from such benchmarking databases as PTB-XL \cite{Wagner+etal:2020}, Georgia (see the 2020 PhysioNet/Computing in Cardiology Challenge \cite{PerezAlday+etal:2020} for details) and CPCS \cite{Feifei+etal:2018}. The distribution of \textit{omeR} by diagnosis and database, as well as by sex and age, is described. In particular, the focus is centered on the PTB-XL results, as it is one of the few databases where BBB are well classified as complete/incomplete, as well as left and/or right. Finally, to facilitate the use of the new tools for the reader, an \textit{app} has been developed that has the registered fragment as input and the new indexes as output. It is available on \url{https://fmmmodel.shinyapps.io/fmmEcg/}.

\section{Methods}
\label{S:2}
\textit{omeR} and \textit{omeS} are defined as functions of the FMM$_{ecg}$ basic parameters.  The FMM$_{ecg}$ model is a particular element of the family of Frequency Modulated M\"obius (FMM) models, which was developed to analyze oscillatory signals and is briefly presented below. 

\subsection{The FMM approach}
Oscillatory signals are defined in the time domain and, without loss of generality, it is assumed that the time points are in $[0,2\pi)$. In any other case, transform the time points $t' \in [t_0,T+t_0]$ by $t=\frac{(t'-t_0)2\pi}{T}$. 

A mathematical term describing an FMM wave is defined as follows:
\begin{equation*}
 W(t,A,\alpha,\beta,\omega) = A \cos(\beta+2\arctan(\omega\tan(\frac{t-\alpha}{2}))),
\end{equation*}
where, $ A \in \Re^+ $ is a scale parameter measuring the wave's amplitude, $\alpha \in [0,2\pi)$ is a location parameter, while $\beta \in [0,2\pi)$ and $\omega \in (0,1]$ are parameters describing the shape. $\beta$ is a skewness parameter that also indicates upward and/or downward peak direction, and $\omega$ measures the width. 

The FMM$_m$ model is defined as an additive signal plus error model, where the signal is defined as a sum of waves as follows: 
\begin{equation*}\label{eq:mod}
  X(t_i)= M + \sum_{J=1}^m W_J(t) + e(t_i); 
\end{equation*}
where $W_J(t)=W(t,A_J,\alpha_J,\beta_J,\omega_J); J=1, \dots ,m$, $M \in \Re $ is an intercept, and the error term, which accounts for the noise, is assumed Gaussian with a common variance. 

In addition, restrictions among the parameters are incorporated depending on the application. The estimators of the parameters are obtained using maximum likelihood. Refer to \cite{Rueda+Rodriguez+Larriba:2021} for the details.

Furthermore, other important parameters of practical use are the peak and trough times for each wave $K$, denoted by $tmax_K$ and $tmin_K$, respectively, and their predicted values denoted by $fmax_K$ and $fmin_K$, respectively, as follows:

\begin{align*}
 tmax_K&=\alpha_K+2 \arctan (\frac{1}{\omega_K}\tan(\frac{-\beta_K}{2})) \\
 tmin_K&=\alpha_K+2 \arctan (\frac{1}{\omega_K}\tan(\frac{\pi-\beta_K}{2})) \\
 fmax_K&= M + \sum_{J=1}^m W_J(tmax_K)\\
 fmin_K&= M + \sum_{J=1}^m W_J(tmin_K)
\end{align*}

and the distances between waves by
\begin{equation*}
 d_{JK}=1-\cos(\alpha_J-\alpha_K); J,K=1,...,m 
\end{equation*}

In the case in which models describing different leads are considered simultaneously, a superscript indicating the lead will be incorporated to the notation.

\subsection{Morphological interpretation of the FMM$_{ecg}$ parameters}
A typical non-pathological ECG signal, such as that shown in Figure \ref{f:typ}, has positive P, R and T waves and non positive Q and S waves, which is well described by the beta FMM parameters: $\beta_J \in arc(2\pi/3,4\pi/3), J=P,R,T$ and  $\beta_J \notin arc(2\pi/3,4\pi/3), J=Q,S$. 
Moreover, the omega and amplitude FMM parameters capture other aspects of the shape of a typical pattern, such as, a prominent and sharp R wave ($A_R$ high and $\omega_R$ low); a prominent and flat T wave ($A_T$ high and $\omega_T$ moderate);  a less prominent and not sharp P wave ($A_P$ low and $\omega_P$ moderate)  and sharp and not prominent Q and S waves ($A_Q,A_S,\omega_Q,\omega_S$ low values). 

Specific values for FMM basic parameters are given for different patterns, including that one, in the supplementary material of \cite{Rueda+Larriba+Lamela:2021}.

\begin{figure}[h!]
	\centering
	\includegraphics[width=1\textwidth]{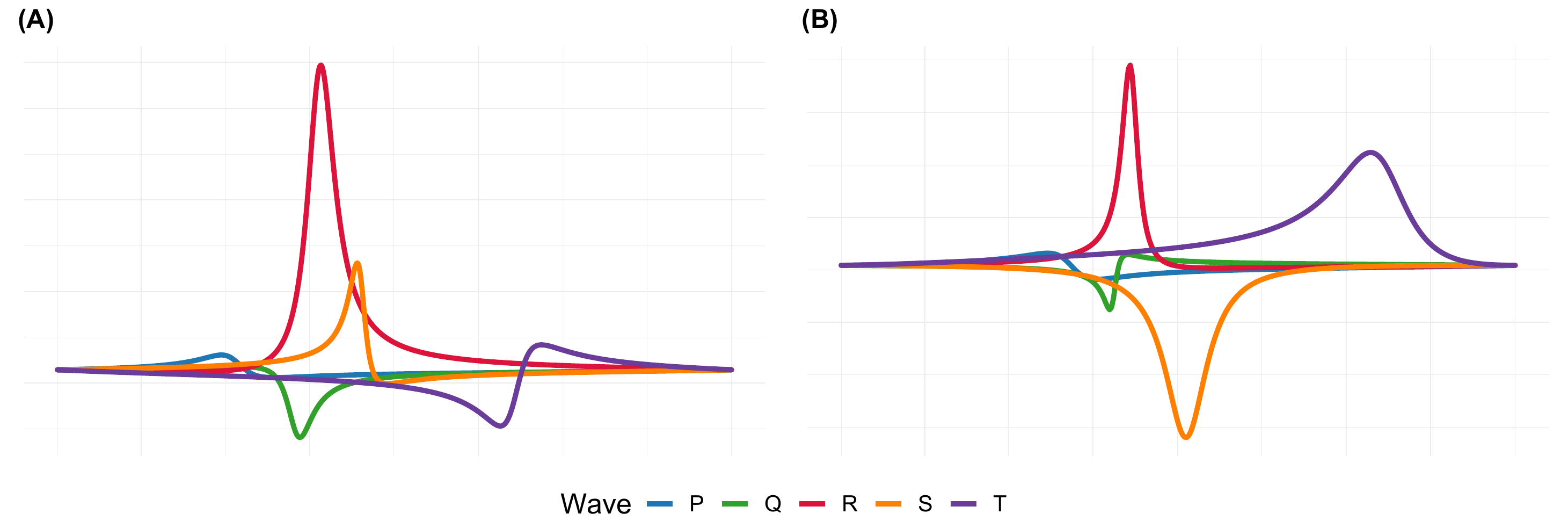}
	\caption{The five waves of FMM$_{ecg}$ fitted to data of complete BBB patients. Left panel (A): data from patient 2087 with CLBBB diagnosis in CPCS database. P wave (blue) $\omega_P=0.110$ and $\beta_P=4.197$; Q wave (green) $\omega_Q=0.070$	and $\beta_Q=5.794$; R wave (red) $\omega_R=0.077$ and $\beta_R=2.930$; S wave (orange) $\omega_S=0.045$ and $\beta_S=3.816$; T wave (purple) $\omega_T=0.092$ and $\beta_T=1.291$. Right panel (B): data from patient 10131 with CRBBB diagnosis in PTB-XL database. P wave (blue) $\omega_P=0.116$ and $\beta_P=4.654$; Q wave (green) $\omega_Q=0.033$ and $\beta_Q=0.892$; R wave (red) $\omega_R=0.045$ and $\beta_R=3.352$; S wave (orange) $\omega_S=0.118$ and $\beta_S=0.042$; T wave (purple) $\omega_T=0.198$ and $\beta_T=3.675$.}
	\label{f:typCLBBB}
\end{figure}

Typical CLBBB and CRBBB patterns can also be described in terms of FMM parameters. In particular, Table \ref{t:diag} includes the description of the criteria for BBB diagnosis enumerated in the introduction. Examples that illustrate the correspondence in Table \ref{t:diag} are included in the \ref{App:A}. Besides, Figure \ref{f:typCLBBB} illustrates the configuration of waves and parameters  that corresponds to the CLBBB and CRBBB signals presented in Figure \ref{f:BBBPatterns}.

\begin{table}[h]
\centering
\begin{tabular}{ll} \hline 
Brief description      & FMM parameter representation \\ \hline
Broad R wave or prolonged R-peak time       & high values of $\omega_R$    \\
Prolonged QRS duration & high values of $d_{QS}$ \\
\textit{M-shaped}, qR-wave & $fmax_Q$ or $fmin_Q$ close to $fmax_R$ \\ 
\textit{M-shaped}, Rs-wave & $fmax_S$ or $fmin_S$ close to $fmax_R$\\
Negative S-wave &  $\beta_S \in arc(5\pi/3,\pi/3) $.\\
Broad S-wave & high values of  $\omega_S$\\
Large S-wave &  high values of $A_S$\\
ST segment depression & high values of $tmin_S$.\\
T-wave inversion &  $\beta_T \in arc(5\pi/3,\pi/3)$.\\
 \hline
\end{tabular}
\caption{Features for BBB diagnostic criteria in leads I, II and V5.}
\label{t:diag}
\end{table}

Alternative parameter configurations to those in Table \ref{t:diag} capture also morphologies with notched R waves. Anyway, notched R waves are often observed as a consequence of $\omega_R$ being high. This property, together with the potential of $\omega_R$ to measure the R peak time and the QRS complex duration, prompts this latter parameter to be an excellent predictor of CLBBB by itself, since many experts associate this disease with the simultaneous combination of this disease occurrence of notches and a prolonged QRS complex. 

\subsection{CLBBB and CBBB diagnostic rules}
In this paper, ECG fragments from leads I, II, and V5 are considered as input. A preprocessing stage is done where basic checks are performed in order to discard high noise signals, correct the trend and perform a simple normalization of the data. Also, a Pan Tompkins algorithm \cite{Pan+Tompkins:1985} is used to locate the QRS complexes used to divide the signal into beats. Moreover, the imputation of missing values has been done. Details of this preprocessing stage are given in the \ref{App:B}. 
  
Now, using an updated algorithm similar to that described in \cite{Rueda+Larriba+Lamela:2021}, the signal is analyzed using the FMM$_{ecg}$. A representative heartbeat is obtained for each patient and lead, using the medians of the parameters. For such a representative heartbeat, let us define: 

\begin{gather*}
\textbf{ CLBBB RULE: \textit{omeR} $> 0.06$} \\
 omeR = \max\limits_{L=\{I, II, V5\}} \{\omega_R^L\}
\end{gather*}

The index developed resembles the Strauss criteria \cite{Strauss+Selvester+Wagner:2011} as it accounts for the QRS duration and R peak time. In fact, \textit{omeR} is the width of the main positive wave in the QRS. More sophisticated rules can be defined using parameters that describe other parts of the morphology, such as the ST-segment depression. However, other diseases, such as RBBB, also has these characteristic patterns, and the specificity decreases while the complexity of the index definition increases. \textit{omeR} itself is a measure of the likelihood for a given patient to have the LBBB disorder.

The typical pattern for a CRBBB is a pronounced negative S wave, which is described with a high value of $\omega_S$ and $\beta_S$ close to $2\pi$.
However, in our databases, more than 5\% of CRBBB patients verify $omeR>0.06$  and would therefore be diagnosed as CLBBB. Then, although more sophisticated rules that increase the accuracy in the sacrifice of simplicity could be defined for specific CRBBB diagnoses, we have opted for a simple rule for a combined CBBB diagnosis that complements the rule defined above.
The combined rule is given as follows:

\begin{gather*}
\textbf{ CBBB RULE: \textit{omeR} $> 0.025$ and \textit{omeS} $> 0.05$}  \\
  omeR = \max\limits_{L=\{I, II, V5\}} \{\omega_R^L\}\\
   omeS = \max\limits_{L=\{I, II, V5\}} \{\omega_S^L  \cdot I_{[\beta_S^L \in  arc(5\pi/3,\pi/3)]}\}
\end{gather*}

where $I_{[ \cdot ]}$ is an indicator taking a value of $1$ if the argument is true and $0$ otherwise.

It is interesting to note that the use of median parameters and the maximum of three values prevent the occasional influence of incorrect parameter estimators that are derived from individual heartbeats and a single lead. These errors may happen due to an incorrect definition of the fragment limits corresponding to a specific heartbeat, signal noise, and a lesser extent, a misidentification of the R wave.

\section{Results}
A total of $106233$ ECG fragments corresponding to $35411$ patients over eighteen and with diverse diagnoses, have been analyzed. The data are from the benchmarking databases PTB-XL (patients with likelihood diagnosis $\geq 80$), Georgia, and CPCS. These databases have been used widely in the literature, to be more precise, in the challenge of Computing in Cardiology 2020 \cite{PerezAlday+etal:2020}. The duration of the ECG signal is quite variable across databases; only the fragment corresponding to the first 14 seconds has been considered for each patient.     
Only 348 patients (less than 1\%), were discarded due to noise artifacts in the signal. The distribution across databases is as follows: 159 (0.86\%) from PTB-XL, 90 (0.88\%) from Georgia and 99 (1.48\%) from CPCS.

Two numerical analyses are presented in this section. The first is dedicated to \textit{omeR}. The distribution of this fundamental marker across diagnosis, sex, and age is explored. The relation of \textit{omeR} with the RR interval, the time elapsed between two successive R waves, is also examined. The second study is the validation of the rules defined above. 

A diagnostic label has been assigned to each patient, using SNOMED CT ontology (\url{bioportal.bioontology.org/ontologies/SNOMEDCT}), as is done for the challenge of Computing in Cardiology 2020 \cite{PerezAlday+etal:2020}.  Thus, patients with a BBB diagnostic are labeled with CLBBB, ILBBB, CRBBB, or IRBBB according to the side (Left/Right) and the degree of the defect (Complete/Incomplete). Note that in the CPCS database, patients with a code LBBB are labeled as CLBBB, as the code for ILBBB is not considered. Furthermore, the label CD is assigned to those with a block diagnostic different to BBB; the NORM label is assigned to patients diagnosed as non-pathological and OTHER to the rest.   

All the patients are classified in one and only one category, even those with several diagnoses (except for a patient with CLBBB and RBBB labelled as LBBB). 

\subsection{\textit{omeR} distribution across diagnostic, age and sex}
In order to consider \textit{omeR} as a \textit{gold standard} to detect anomalous ECG is necessary to know reference values and normal ranges across diagnostics, sex and age. 

Median values and the percentage interval for \textit{omeR} are given for patients in the PTB-XL, Georgia and CPCS databases, across diagnostics in Table \ref{t:omeR2}. 

\begin{table}[h!]
\centering
\begin{tabular}{llrccc} \hline 
Database & Diagnosis & N       & $P_5$ & $P_{50}$ & $P_{95}$ \\ \hline 
PTB-XL   & CLBBB     & 503     & 0.080 & 0.148    & 0.243    \\ 
         & BBB       & 2168    & 0.028 & 0.047    & 0.190     \\ 
         & CD        & 2529    & 0.029 & 0.048    & 0.109    \\   
         & OTHER     & 5275    & 0.026 & 0.043    & 0.088    \\ 
         & NORM      & 8310    & 0.026 & 0.036    & 0.052    \\
\hline
Georgia  & CLBBB     & 208     & 0.063 & 0.137    & 0.246    \\ 
         & BBB       & 1170    & 0.027 & 0.052    & 0.180    \\ 
         & CD        & 1024    & 0.026 & 0.045    & 0.099    \\  
         & OTHER     & 6253    & 0.025 & 0.044    & 0.084    \\   
         & NORM      & 1723    & 0.027 & 0.038    & 0.058    \\ 
\hline
CPCS     & CLBBB     & 226     & 0.057 & 0.133    & 0.236    \\ 
         & BBB       & 2044    & 0.028 & 0.043    & 0.144    \\ 
         & CD        & 688     & 0.026 & 0.038    & 0.066    \\   
         & OTHER     & 3067    & 0.029 & 0.045    & 0.089    \\  
         & NORM      & 812     & 0.029 & 0.040    & 0.055    \\ 

\hline 
\end{tabular}
\caption{Percentile range and median for \textit{omeR} across diagnostic in the PTB-XL, Georgia and CPCS databases.}
\label{t:omeR2}
\end{table}

Moreover, we have claimed in the introduction that an important property of \textit{omeR} is that it is independent of sex. Table \ref{t:RR} shows percentile ranges and median values for the RR interval and \textit{omeR} across databases, age and sex. Only NORM patients are considered, as spurious associations could be found if other diagnoses were included because they are related to \textit{omeR} alterations.

The numbers in Table \ref{t:RR} show that the distribution of \textit{omeR} does not depend on sex or age, while the distribution of the RR interval does. Increasing values of the RR interval medians are obtained with increasing age, and for males against females, as is well documented in the literature \cite{Dhingra+etal:2006, Badheka+etal:2013}.
Moreover, it is also shown that the \textit{omeR} distributions are much more homogeneous across databases than those of the RR interval.

\begin{table}[h!]
\centering
\begin{tabular}{llrrrrccc} \hline
\multicolumn{2}{c}{ } & & \multicolumn{3}{c}{RR interval} & \multicolumn{3}{c}{ \textit{omeR}} \\ \cline{4-9}
         &            & N    & $P_5$ & $P_{50}$ & $P_{95}$ & $P_5$ & $P_{50}$ & $P_{95}$ \\ \hline
PTB-XL   & Male       & 3774 & 648  & 899    & 1183    & 0.026  & 0.036  & 0.052 \\ 
         & Female     & 4536 & 620     & 856    & 1125    & 0.026  & 0.036  & 0.051 \\ \cline{2-9}
         & $< 25$     & 461  & 636   & 864    & 1172  & 0.025  & 0.035  & 0.047 \\   
         & $[25, 50]$ & 3104 & 634     & 872    & 1146    & 0.026  & 0.036  & 0.051 \\ 
         & $[51, 70]$ & 3495 & 645     & 890    & 1157  & 0.026  & 0.035  & 0.051 \\
         & $>70$      & 1250 & 592  & 842    & 1132 & 0.027  & 0.037  & 0.054\\
\hline
Georgia  & Male       & 792  & 618     & 819    & 988     & 0.028	 & 0.038   & 0.060  \\ 
         & Female     & 931  & 628	   & 809	& 983     & 0.027	 & 0.038   & 0.056 \\ \cline{2-9}
         & $< 25$     & 50   & 634     & 771	& 946  & 0.028	 & 0.040    & 0.061 \\   
         & $[25, 50]$ & 605  & 621     & 800	& 984   & 0.028	 & 0.039   & 0.057 \\ 
         & $[51, 70]$ & 791  & 620     & 822	& 988     & 0.027	 & 0.037   & 0.058 \\
         & $>70$      & 277  & 636   & 834	& 986   & 0.027	 & 0.037  & 0.058 \\
\hline
CPCS     & Male       & 322  & 640   & 831	& 1008 & 0.029	 & 0.040    & 0.054 \\ 
         & Female     & 490  & 629     & 809	& 970   & 0.030 	 & 0.040   & 0.055 \\ \cline{2-9}
         & $< 25$     & 52   & 657   & 800	& 944    & 0.035	& 0.042	  & 0.054 \\   
         & $[25, 50]$ & 452  & 632   & 806	& 969 & 0.029    & 0.041   & 0.055 \\ 
         & $[51, 70]$ & 244  & 654   & 846	& 1015 & 0.029	& 0.039   & 0.054 \\
         & $>70$      & 64   & 618   & 857	& 1027 & 0.028    & 0.038   & 0.053\\
\hline 
\end{tabular}
\caption{RR interval and \textit{omeR} percentile ranges and median for NORM patients across gender and age groups in the PTB-XL, Georgia, and CPCS databases.}
\label{t:RR}
\end{table}

\subsection{Rule performance in the PTB-XL, Georgia and CPCS databases}
Standard measures of test validation, such as sensitivity (SE) or specificity (SP), have been calculated using  both rules for different types of patients in Table \ref{t:omeRrule} and \ref{t:omeSrule}, respectively. 

The target of the rules is patients with CLBBB or CRBBB. However, the differentiation between complete or incomplete blocks is not homogeneous across the databases, partially explaining the differences in sensitivity between them. Moreover, the distribution of patients across other diagnoses also explains the differences in specificity for the category ALL.
 
The PTB-XL(1) database is defined as the sub-base containing PTB-XL patients with a single diagnosis. The PTB-XL(1) is a good reference, as patients have a unique diagnostic and it discriminates between complete or incomplete blocks. For this database, the sensitivity of CLBBB (CBBB) rules is 99\% (98\%), the specificity is 99\% (90\%) for NORM patients, and 96\% (87\%) for patients without an LBBB (CBBB) diagnostic. Sensitivity and specificity values are very similar for PTB-XL.
  
Moreover, in the Georgia and CPCS databases, the slightly lower sensitivity values obtained for the CLBBB rule, 96\% and 93\%, respectively, compared with those of PTB-XL, may reflect the fact that the degree of injury is not well differentiated in these databases. Specifically, some patients labeled as LBBB could be ILBBB. Sensitivity values for the CBBB rule are similar to that of PTB-XL.

On the other hand, the specificity is very high, especially for NORM patients and the CLBBB rule, where the values range from 96\% in Georgia to 99\% in PTB-XL.

\begin{table}
\centering
\begin{tabular}{lcccccc} \hline
           & N      & SE(CLBBB) & SE(ILBBB) & SE(LBBB) & SP(NORM) & SP(ALL) \\ \cline{2-7}
PTB-XL(1)  & 12393  & 99\%      & 67\%      & 99\%     & 99\%     & 96\%     \\
PTB-XL     & 18282  & 98\%      & 88\%      & 97\%     & 99\%     & 88\%     \\
Georgia    & 10170  & 96\%      & 68\%      & 88\%     & 96\%     & 84\%     \\ 
CPCS       & 6611   & 93\%      & ---       & 93\%     & 98\%     & 86\%     \\ 
\hline 
\end{tabular}
\caption{CLBBB rule: Sensitivity (SE) and specificity (SP) across subgroups of patients by diagnostic. The ALL category includes all the patients except those with LBBB labels.}
\label{t:omeRrule}
\end{table}

\begin{table}
\centering
\begin{tabular}{lcccccc} \hline
           & N      & SE(CBBB)  & SE(IBBB)  & SE(BBB)  & SP(NORM) & SP(ALL) \\ \cline{2-7}
PTB-XL(1)  & 12393  & 98\%      & 27\%      & 67\%     & 90\%     & 87\%     \\
PTB-XL     & 18282  & 96\%      & 34\%      & 74\%     & 90\%     & 79\%     \\
Georgia    & 10170  & 96\%      & 53\%      & 72\%     & 87\%     & 72\%     \\ 
CPCS       & 6611   & 95\%      & ---       & 74\%     & 92\%     & 77\%     \\ 
\hline 
\end{tabular}
\caption{CBBB rule: Sensitivity (SE) and specificity (SP) across subgroups of patients by diagnostic. The ALL category includes all the patients except those with BBB labels.}
\label{t:omeSrule}
\end{table}

\subsection{FMM$_{ecg}$ \textit{app}}
The \textit{app} is freely available on \url{https://fmmmodel.shinyapps.io/fmmEcg/}. The instructions for use are very simple. The \textit{app} requires the ECG fragment from either I, II, or V5 leads, including one or multiple heartbeats, recorded, as input. The output includes: plots of the observed and predicted values, a plot of the FMM median waves, tables with the median values of the estimated FMM parameters, the $omeR$ and $omeS$ markers, and the BBB diagnosis. Note that the values of $omeR$ and $omeS$ calculated in the \textit{app} are obtained from a single lead. When that information on multiple leads is available, the values of the markers can be obtained for each lead independently with the \textit{app}, and then the maximum of the marker values is calculated by hand.

In the \ref{App:C} provides further details about the \textit{app}, including multiple-use examples.

\section{Discussion}
In this paper, new ECG markers are proposed, defined using the FMM$_{ecg}$ delineator. In the main place, \textit{omeR}, which is related to the QRS duration, a widely used index that is relevant to know. In particular,  the literature has shown that a large QRS duration is a characteristic associated to different diseases such as ischemic cardiomyopathy \cite{Iuliano+etal:2002}, myocardial infarction \cite{Fosbol+etal:2007} or sudden death \cite{Teodorescu+etal:2011}.

Compared with the latter index, \textit{omeR} has a more precise meaning, is measured in a normalized scale, is independent of sex, and does not depend on the measuring device or the researcher, as it is a parameter of a statistical model that is estimated by maximum likelihood.
Normal ranges for \textit{omeR} have been provided, as well as ranges across diagnostics, sex, and age. Relevant differences have been observed only across diagnostics. Then, the association of the QRS duration with age and sex, which is documented in the literature \cite{Dhingra+etal:2006, Badheka+etal:2013}, can be explained by the association of these three variables with the RR interval length. 

Furthermore, the results in this and previous works \cite{Rueda+Larriba+Peddada:2019, Rueda+Larriba+Lamela:2021, Rueda+Rodriguez+Larriba:2021}, have shown that the FMM$_{ecg}$ parameters are estimated with high accuracy, and they are consistent and robust. In particular, \textit{omeR} because it is based on the main median parameter of the most prominent wave and is less affected by noise. 
Besides, the FMM$_{ecg}$ parameters have much potential for detecting causes and diagnoses derived from a prolonged QRS as they describe the morphology of the QRS complex in an exact way, with four parameters describing the shape and length of each of the waves Q, R, or S. In particular, while a wide R wave is associated with left blocks, a prominent and negative S wave describes the typical morphology of a right block. In both cases, the QRS duration is large, and \textit{omeR} and \textit{omeS} differentiate these two diseases automatically from other patterns.
 
Despite their simplicity, the ability of the two classifiers proposed in this paper to correctly identify a BBB is highly satisfactory. The results are particularly good for the CLBBB rule, which confirms the potential of \textit{omeR} in diagnosis. Specifically, the sensitivity and specificity values obtained for the CLBBB classification are similar or even higher than those obtained from approaches proposed by other authors, such as \cite{Moody+Mark:2001,Hao+etal:2018,Allami+etal:2016,MartinYebra+Martinez:2019,Yang+etal:2020}. Nevertheless, comparisons with other rules are not fair as none of them has such universal character or they are not possible to be applied automatically.

It is noteworthy that the rules can be adapted to the information available. On the one hand, they can be used with data from only one or two leads, and on the other hand, when the \textit{omeR} and \textit{omeR}  baseline values are known in advance for a given patient, the thresholds in the rules can be redefined to be  percentages of these values.

It is also worth pointing out that the databases used to validate the rules are not explicitly designed for the problem at hand. They are databases collected and labelled for very different purposes and consequently exhibit an entirely different distribution of patients across diagnostics, which may explain the difference in accuracy values across databases.  However, they do coincide in showing much better results in identifying complete left blocks compared to the complete right or the incomplete ones. Nevertheless, \textit{omeR} and \textit{omeS} could provide an opportunity to study the evolution of the degree of BBB and even prevent more severe blocks in the future. In fact, according to some authors, ILBBB (IRBBB) might be a precursor of CLBBB (CRBBB) \cite{Surawicz+etal:2009,Senesael+etal:2020}. 

The usefulness of \textit{omeR}, \textit{omeS} and other markers derived from the FMM$_{ecg}$ basic parameters is promising and vast and goes much further than the diagnosis of BBB and anything we can say in this paper. In particular, these markers are much more easily registered than serial ECG, so current values are compared with those obtained at other moments in time that can run from one hour to many years before. On the other hand, the automatic registration of the markers, not necessarily supervised by an expert, facilitates many tasks. For instance, monitoring \textit{omeR} after a trascatheter aortic valve replacement could be essential to identifying an LBBB that persists more than 72 hours and assist in prophylactic decision making \cite{Tan+etal:2020}. Alternatively, it could also identify the presence of a new BBB in patients with an acute myocardial infarction, in which pacemaker insertion may not be not beneficial \cite{Epstein+etal:2008}.

Finally, the derivation of consistent markers for the automatic interpretation of the ECG would reduce the observed differences in the interpretation of the electrocardiographic abnormalities among health professionals, the first cause being a manual reading of the ECGs. Unifying the criteria would help improve the competence of non-cardiologist physicians and achieve a better and automatic diagnosis.

Several lines of work for future research emerge from this paper. On the one hand, the definition of rules for the diagnosis of other diseases. Specifically, the automatic diagnosis of myocardial infarction, using FMM$_{ecg}$ parameters, will be our immediate challenge. On the other hand, the simultaneous estimation of signals from different leads using multivariate models would reduce the already rare cases of the incorrect identification of waves or anomalous parameter estimators.

\section*{Acknowledgements}
The authors gratefully acknowledge the financial support received by
the Spanish Ministerio de Ciencia e Innovaci\'on [PID2019-106363RB-I00] to CR, IF, YL as well the Call for predoctoral contracts of the UVa 2020 to AR-C.

\section*{Author contributions}
CR: Conceived aims, theoretical proposal, conceptual design, wrote the manuscript. CR,IF: revised specialized bibliograhy and propose LBBB criteria.  CR, YL design the preprocesing stage. YL, AR-C, CC: developed the computational code and the \textit{app}.
All the authors approved the manuscript. 

\section*{Declaration of competing interest}
The authors declare that they have no known competing financial interests or personal relationships that could have appeared to influence the work reported in this paper.

\newpage
\appendix
\section{FMM$_{ecg}$ Examples for BBB Features} \label{App:A}
\setcounter{figure}{0}   
This appendix shows examples that illustrate how FMM$_{ecg}$ parameters describe BBB features as those given in Table \ref{t:diag} of the main document. Row 1 in Figure \ref{f:ex} corresponds to a typical (non-pathological) pattern simulated using the median FMM$_{ecg}$ parameters obtained from the analysis of NORM patients in PTB-XL.

The rows 2-5 show patterns using alternative parameter configurations, which are derived from that of row 1. The inner tables in each of the plots in Figure \ref{f:ex} show the specific FMM$_{ecg}$ parameter configuration that differs from the typical pattern.

Specifically, $\omega_R$ is related to the width of the R wave, as the plots in row 2 show. Besides, row 3 illustrates that a higher distance between $\alpha_Q$ and $\alpha_S,$ is related with a prolonged QRS duration, and also that changes in $A_Q$ and $\beta_Q$ (resp. $A_S$ and $\beta_S$) yield a qR (resp. Rs) notched pattern. Parameter configuration related to S wave is illustrated in row 4, together with a specific configuration for ST segment depression. Finally, T wave inversions are observed in row 5. 

\begin{figure}[H]
\centering
\includegraphics[width=0.9\textwidth]{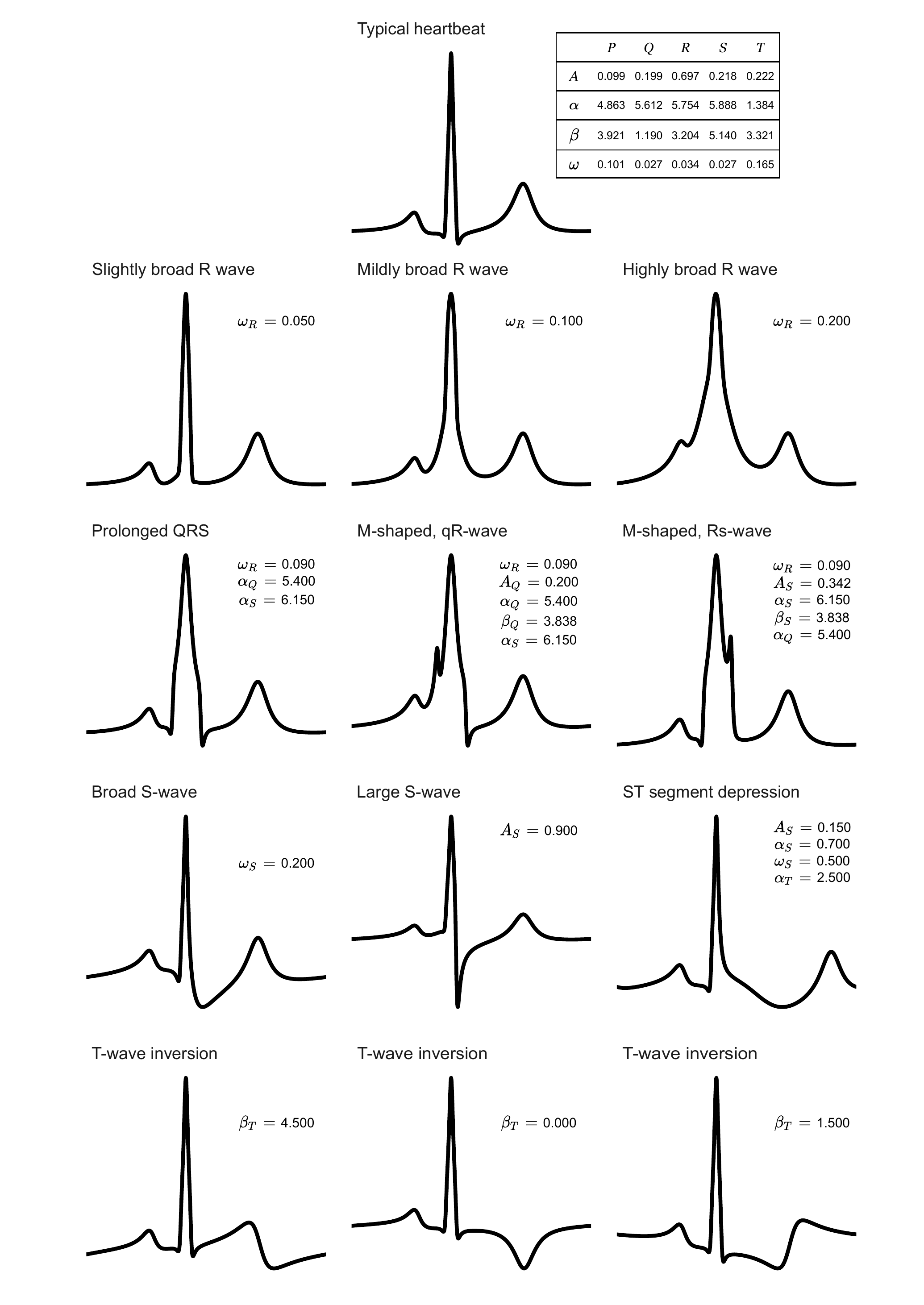}
\caption{Simulated examples to illustrate BBB diagnostic features showed in Table \ref{t:diag} of the main text}
\label{f:ex}
\end{figure}

\section{Data preprocessing} \label{App:B}
\setcounter{figure}{0}
ECG data preprocessing is required to remove high noise signals, or other motion artifacts in the recording, to normalize the signal and to correct the trend on ECG heartbeats. Conventionally, ECG data come from multiple leads and their preprocessing is based on the following stages: ECG denoising, QRS detection, ECG segmentation, scaling  and trend correction \cite{Aspuru+etal:2019,Tseng+etal:2016}. Most of the preprocessing procedures proposed in the literature cover these stages. A brief discussion of that is given below. 

Most of noise suppression techniques are often based on band-pass filterings, such as the Butterworth filter \cite{Friganovic+etal:2018, Pandey+Tiwari:2014}. They highly rely on a fixed cut-off frequency and cannot track the changing characteristics of the time-varying ECG signal. As a result, the ST segment or the QRS complex may be notably distorted \cite{Sun+etal:2002}.  

Regarding QRS detection, several methods have prolifered in the last decade \cite{Chen+Chuang:2017,Elgendi+etal:2014}. However, Pan-Tompkins algorithm \cite{Pan+Tompkins:1985}, reported in 1985, is considered as the benchmark in QRS detection. Once these complexes are set, ECG heartbeats must be delimeted, existing a wide range of strategies to cover this goal. Among others, there are those based on P onset and T offset locations \cite{Sahoo+etal:2018}, although these marks are not precisely defined in practice. While in others \cite{Rueda+Larriba+Lamela:2021,Tseng+Zeng+etal:2016}, the cutting is based on the length of the RR intervals, i.e. the time elapsed between two successive QRS detections. 

ECG segmentation often displays a high degree of uncertainty, due to the noise in the device, missing sections of the ECG signals, or because the signal contains a transient rise of signal amplitude \cite{Nurmaini+etal:2021}. In such cases, Pan Tompkings usually fails resulting in missed peaks or false detections \cite{Arefin+etal:2015,Mondelo+etal:2017}. Therefore, an ECG segmentation revision is desirable. Most of these procedures are based on the RR interval properties such as length or density \cite{Sadhukhan+Mitra:2012,Tseng+Zeng+etal:2016}.

The ECG signal normalization is commonly based on linearly scaling methods \cite{Yang+etal:2013,Liang+etal:2020}. While polynomial regression is widely extended for ECG detrending \cite{Friganovic+etal:2018,Aspuru+etal:2019}. Finally, ECG fragment distortions, caused by sudden body movements, or other noise artifacts, are usually removed if the heartbeat's amplitude is relatively high with regard to that observed in the signal \cite{Chu+Delp:1990,Sornmo+Laguna:2006}. 

\begin{figure}[H]
\includegraphics[width=1\textwidth]{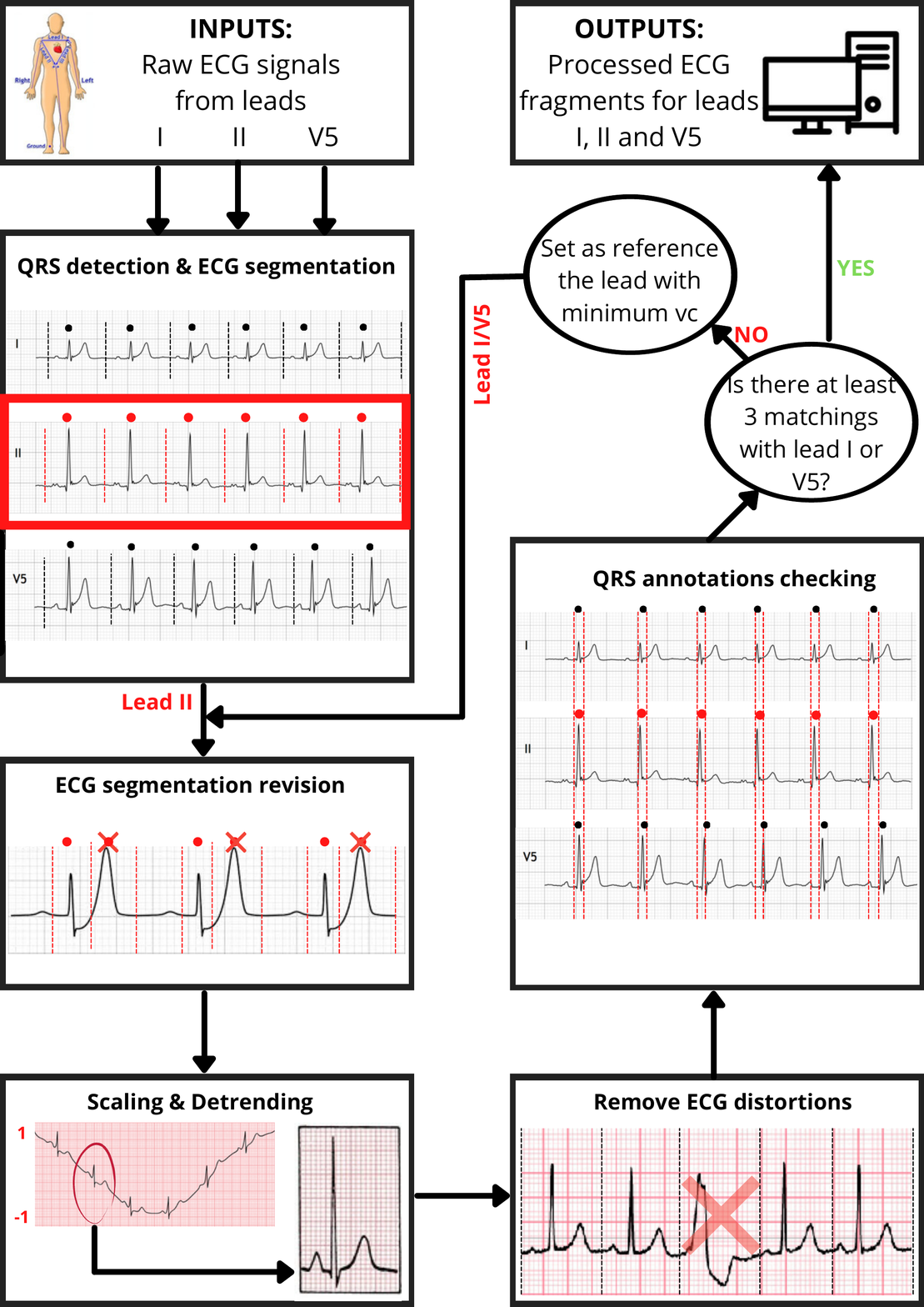}
      \caption{Outline of the three-lead preprocessing algorithm}
          \label{f:alg}
\end{figure}

A relatively simple data preprocessing algorithm is defined here. First, noise suppression is dismissed in this work since FMM$_{ecg}$ is robust against noise artifacts \cite{Rueda+Larriba+Lamela:2021}. For QRS detection, Pan Tompkins algorithm is applied on raw ECG data, from one or three leads. ECG heartbeats are delimited based on the RR intervals. ECG signals are normalized using a linearly scaling and heartbeat detrending is based on linear regression. The details of the preprocessing are described below. An overview of this algorithm is illustrated in Figure \ref{f:alg}.

\begin{flushleft}
\textbf{Three-lead algorithm:} The raw ECG signals from the leads I, II and V5 are the inputs. The five steps of the algorithm are:
\end{flushleft}
\begin{enumerate}
\item \textbf{QRS detection and ECG segmentation:} Raw ECG signals from leads I, II and V5 are independently analyzed with Pan Tompkins. Then, ECG hearbeats are delimited from the QRS annotations ($t^{QRS}$) and RR interval lengths as follows: [$t^{QRS}-40\%RR,$ $t^{QRS}+60\%RR$], see \cite{Rueda+Larriba+Lamela:2021} for details. Lead II is used as reference \cite{Meek+Morris:2002,Mondejar+etal:2019}. 

\item \textbf{ECG segmentation revision:} Median RR length is used to identify and remove too short/long heartbeats and those with distant R peaks or misplaced $t^{QRS}$ locations, following the ideas given in \cite{Rezk+etal:2011,Sadhukhan+Mitra:2012,Rueda+Larriba+Lamela:2021}. 

\item\textbf{Scaling and detrending:} ECG recording is scaled into [-1,1] using min-max normalization as in \cite{Liang+etal:2020}. Next, as noted in \cite{Sun+etal:2005,Arefin+etal:2015,Liang+etal:2020}, those heartbeats with significant trends or remarked differences at the heartbeat's boundaries, are detrended by using linear regression \cite{Mitov:1998,Sharma+Sharma:2017,Kontaxis+etal:2021}.

\item\textbf{Remove ECG distortions:} Heartbeats whose amplitudes are considerably
larger with regard to the QRS amplitudes in the signal are removed as done in \cite{Chu+Delp:1990,Sornmo+Laguna:2006}. 

\item  \textbf{QRS annotations checking}: The QRS annotations from two leads are considered to be in agreement if their distance is less than that corresponding to 4\% frequency (Hz). If no two leads coincide in at least three QRS annotations, the patient is discarded. Otherwise, in the case the reference lead does not match with any of the others in at least three QRS annotations, the reference derivation is changed to that with lower RR length variation coefficient (vc) among I and V5. In this latter case the algorithm goes back to \textit{Step2}. Otherwise, the preprocessing has finished.
\end{enumerate}

For each lead, the outputs are ECG fragments with valid or invalid (removed) heartbeats. Any patient with less than three valid heartbeats in a given lead is discarded. 

Removed heartbeats are imputed after the FMM$_{ecg}$ analysis. The imputed values are the median FMM$_{ecg}$ parameter values obtained from the valid heartbeats in the corresponding fragment.

\begin{flushleft}
\textbf{Single-lead algorithm}: The raw ECG signals one leads are the inputs. The algorithm given above must be adapted as follows. \textit{Step 1} is conducted just focusing on the input lead. Then, \textit{Steps 2-4} are conducted similarly. \textit{Step 5} reduced to discard patients with less than three valid heartbeats.
\end{flushleft}

\section{FMM$_{ecg}$ Analysis \textit{app}} \label{App:C}
\setcounter{figure}{0}
\subsection{Implementation}
The application has been developed in the programming language \texttt{R}, using the package \texttt{shiny} \cite{shiny}. Shiny applications are mainly composed of two modules: the User Interface (UI), designed with \texttt{HTML} widgets and \texttt{CSS} elements, and the Server, responsible for the computational tasks implemented in \texttt{R}.

The UI module makes use of several \texttt{R} packages to create a complete and flexible interface. In particular, the dashboard appearance is given by \texttt{shinydashboard} \cite{shinydashboard} and \texttt{flexdashboard} \cite{flexdashboard} packages. The dashboard has been implemented with the packages \texttt{shinyWidgets} \cite{shinyWidgets}, \texttt{shinyjs} \cite{shinyjs}, and \texttt{ggplot2} \cite{ggplot2}, which provide a collection of control elements, \texttt{JavaScript} interactions, and a visualization engine, respectively.

The data analysis using the FMM approach is implemented using the \texttt{FMM} package \cite{Fernandez+etal:2021, FMMpackage}. Detailed information on the model specifications and the estimation algorithm are given in \cite{Rueda+Larriba+Lamela:2021}.

\subsection{Application structure}
\begin{figure}[H]
\centering
\includegraphics[width=\textwidth]{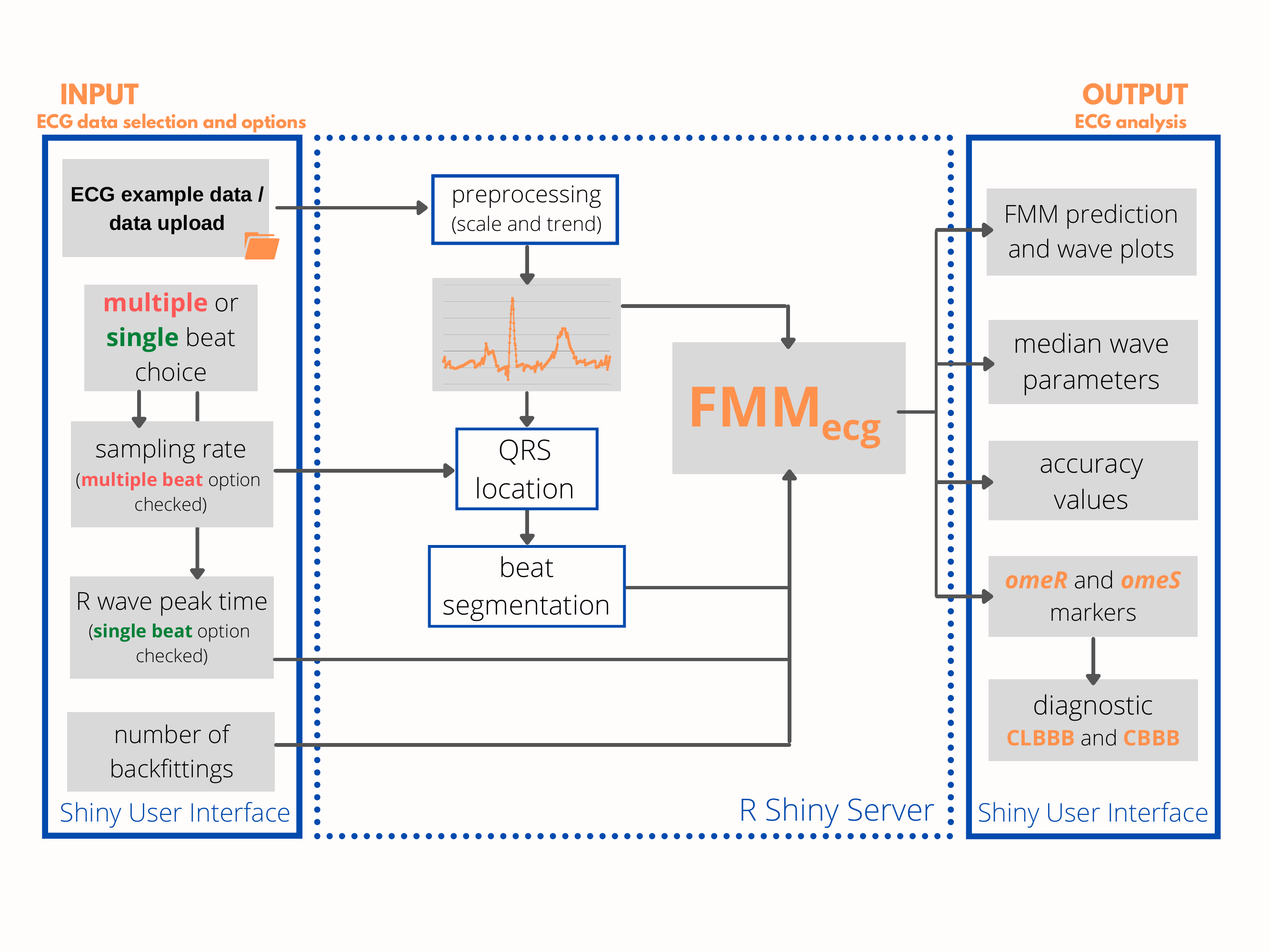}
\caption{FMM$_{ecg}$ Analysis \textit{app} workflow chart}
\label{apptfg}
\end{figure}

An overview of the \textit{app} workflow is shown in Figure \ref{apptfg}. Flexibility is granted to the \textit{app} by its different input elements, which are the following:

\begin{itemize}
    \item \textbf{Main input}. An uploaded tabular data file, which can either have a single beat or multiple recorded in one of the I, II, or V5 leads. It must contain a single column with a column header name and be one of the following extensions: .csv, .xls, or .xlsx. 
    
    \item \textbf{Data options}. User may indicate whether the data is composed by a single beat or by multiple beats. The user must insert the observation number of the QRS annotation, in the case of single beat analysis, and the sampling rate in Hz, for multiple beat analysis. 
    
    \item \textbf{Algorithm options: number of backfittings}. More backfitting iterations imply a better prediction at the cost of a higher computational time.
\end{itemize}

Once the previous options have been correctly specified, the FMM$_{ecg}$ analysis can be started by pressing the prediction button. The following outputs are given:

\begin{itemize}
    \item \textbf{The predicted FMM signal plotted along with the input data.}
     \item  \textbf{Accuracy measures: $var_J; J=P,Q,R,S,T$} are measures accounting for the percentage of the variability explained for each wave relative to the previous fitted waves. Let J be the k-th wave fitted, then:
     
   \begin{equation*}\label{eq:PV}
var_J = R^2_{1,...,k}- R^2_{1,...,k-1},
\end{equation*}
where $R^2_{1,...,k}$, is the proportion of variance explained by the FMM model defined by the first k waves, out to the total variance.

    \item \textbf{Individual waves plot.} Plot of estimated $W_J(t),\,J\in\{P,Q,R,S,T\}$, $t \in (0,2\pi]$. 
    
    \item \textbf{Median wave parameters.} Median estimated values of $ A_J, \alpha_J, \beta_J, \omega_J$, $J\in\{P,Q,R,$ $S,T\}$.
    
    \item \textbf{\textit{omeR} and \textit{omeS} estimated markers and diagnostic button}. The diagnostic is made using the \textit{omeR} and \textit{omeS} rules to detect CLBBB and CBBB pathologies, respectively.
    
\end{itemize}

Predicted signal and individual waves plots are shown simultaneously when data corresponds to a single beat. In the case of analysing multiple beat signal, the \textit{app} provides a button to shift between the two plots.

For demonstration purposes, example data is also provided. In particular, a single healthy beat, a single beat with CLBBB and a multiple beat signal with CBBB.

\subsection{\textit{app} examples}
Three examples from PTB-XL database have been included in the \textit{app}. The first example (healthy) is the ECG signal recorded from patient 14, lead II, the second example (CLBBB) is the ECG signal recorded from patient 796, lead I, and the third example (CBBB) is a multiple beat signal recorded from patient 10131, lead V5. Upon initialization, the \textit{app} analyzes the first example. The \textit{app} automatically provides the data options if example data is analyzed, which for this case are the `single beat' option and  68 for the QRS annotation. Lastly, the user can choose the number of backfittings of the FMM algorithm, which can be initiated by pressing the prediction button. The \textit{app} shows the predicted signal and individual waves plots, the high prediction accuracy, the estimated parameter values for each wave, and the \textit{omeR} and \textit{omeS} markers. By clicking the diagnostic button, the pathology rules are evaluated. In the first example the patient is diagnosed as healthy.

Similarly, in the second example, the data inputs are analyzed with the options `single beat'  and 70 for the QRS annotation. While for the third example, the options `multiple beats' and $200$ Hz for the sampling rate must be provided. Also in these two cases, pathologies are diagnosed, in particular, CLBBB in the second example and CBBB in the third.   

Users can upload their own ECG fragment to be analyzed through the FMM approach with the corresponding options/buttons on the left side of the screen. After uploading the data,  whether the ECG fragment contains a single or multiple beats must be indicated and also the QRS annotation, in the former case, or the sampling rate, in the latter case. Finally, the user should choose the number of backfittings with the correponding button.



\bibliographystyle{model1-num-names}
\bibliography{ref_BBB.bib}







\end{document}